\documentclass{article}
\usepackage{spconf,amsmath,amssymb,graphicx}
\usepackage[T1]{fontenc}
\usepackage[utf8]{inputenc}
\usepackage[table]{xcolor}
\usepackage{booktabs,multirow,tabularx,comment,enumitem}
\usepackage[hidelinks]{hyperref}
\usepackage{microtype,etoolbox}

\AtBeginEnvironment{table}{}
\AtBeginEnvironment{table*}{}
\AtBeginEnvironment{thebibliography}{\setlength{\itemsep}{1pt}\setlength{\parskip}{0pt}}
\apptocmd{\thebibliography}{\setlength{\itemsep}{0pt}\setlength{\parsep}{0pt}}{}{}
\definecolor{indigo}{RGB}{75,0,130}

\setlist[itemize]{leftmargin=1.4em,itemsep=1pt,parsep=0pt,topsep=3pt}
\title{ML-ITW: A Multilingual In-the-Wild Benchmark \\
	for Speech Deepfake Detection}
\name{\shortstack{Daixian Li$^{*}$, Jun Xue$^{*}$, Zhuolin Yi$^{}$, Yanzhen Ren$^{\dagger}$, 
Yihuan Huang$^{}$, Guanxiang Feng$^{}$, Yi Chai$^{}$}\thanks{$^{*}$These authors contributed equally. $^{\dagger}$indicates the corresponding author.}}
\address{\fontsize{10}{12}\selectfont\shortstack{ School of Cyber Science and Engineering, Wuhan University\\
\texttt{hiimchlo1@whu.edu.cn, junxue@whu.edu.cn, renyz@whu.edu.cn}}}
\begin{document}
\ninept
\maketitle
\begin{abstract}
    Recent advances in speech synthesis and voice conversion have greatly improved the naturalness and authenticity of generated audio. Meanwhile, evolving encoding, compression, and transmission mechanisms on social media platforms further obscure deepfake artifacts. These factors complicate reliable detection in real-world environments, underscoring the need for representative evaluation benchmarks. To this end, we introduce ML-ITW (Multilingual In-The-Wild), a multilingual dataset covering 14 languages, seven major platforms, and 180 public figures, totaling 28.39 hours of audio. We evaluate three detection paradigms: end-to-end neural models, self-supervised feature-based (SSL) methods, and audio large language models (Audio LLMs). Experimental results reveal significant performance degradation across diverse languages and real-world acoustic conditions, highlighting the limited generalization ability of existing detectors in practical scenarios. The ML-ITW dataset is publicly available\footnote{https://huggingface.co/datasets/chlo1/ML-ITW}.
\end{abstract}

\section{Introduction}

Recent advances in neural speech synthesis and voice conversion enable generated audio to closely mimic human timbre, prosody, and semantic coherence, bringing both practical benefits and security risks such as identity impersonation, misinformation and large-scale manipulation of public opinion. To counter such threats, speech deepfake detection has been extensively studied and benchmarked. Standardized evaluation campaigns such as the ASVspoof series \cite{wu15e_interspeech, wang2020asvspoof, yamagishi2021asvspoof2021acceleratingprogress, wang24_asvspoof} have driven rapid methodological progress, fostering diverse detection paradigms, including end-to-end architectures operating on spectral or waveform inputs (e.g., LCNN \cite{lavrentyeva2019stc}, RawNet2 \cite{tak2021end}, AASIST \cite{jung2022aasist}), self-supervised hybrids leveraging pretrained encoders (e.g., XLSR+\allowbreak AASIST \cite{tak2022automatic}, ML\_SSLFG \cite{tran2025multi}), and emerging audio large language models (e.g., ALLM4ADD \cite{gu2025allm4add}, HoliAntiSpoof \cite{xu2026holiantispoof}). Despite strong performance on controlled benchmarks, their robustness under realistic deployment conditions remains to be further investigated.

In real-world scenarios, audio distributed through social media platforms undergoes encoding, compression, transcoding, and redistribution, with platform-specific processing strategies introducing distinct artifacts. These transformations alter spectral and statistical characteristics, leading to cross-platform distribution shifts that can degrade detection performance. Therefore, evaluating models under multi-platform propagation conditions is critical to understand their practical generalization ability.

\begin{figure}[t!]
  \centering
  \includegraphics[width=\linewidth]{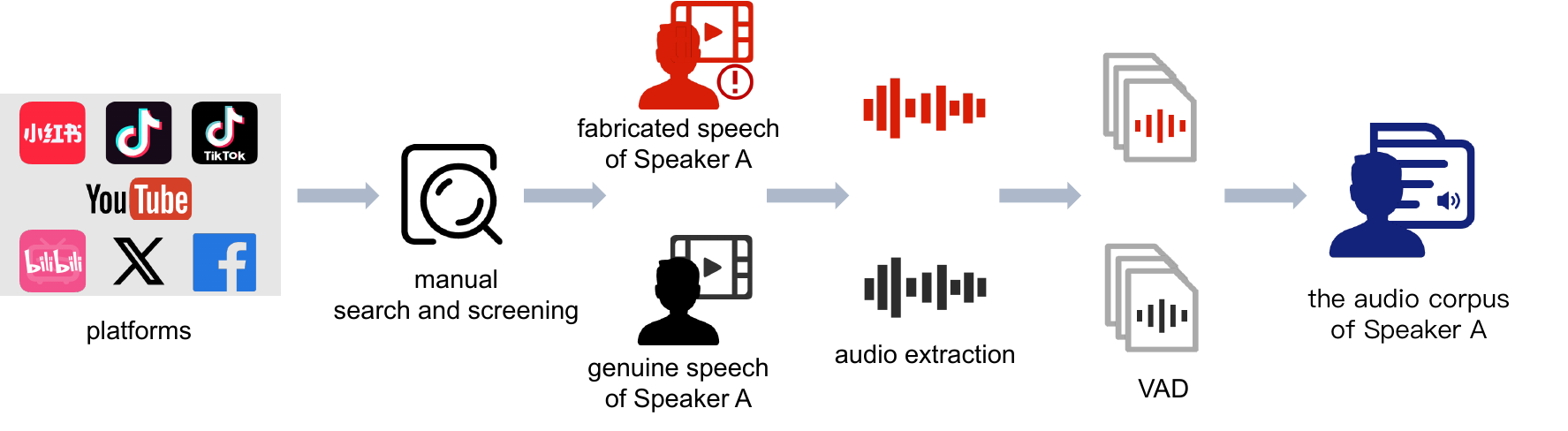}
  \caption{Data collection pipeline for a specific speaker in the ML-ITW dataset.}
  \label{fig:dataset_construction}
  
\end{figure}

Although numerous datasets have been proposed to support speech deepfake research, most are constructed in controlled settings with relatively homogeneous sources and processing pipelines. ASVspoof2019, ASVspoof5, WaveFake \cite{frank2021wavefake}, ADD \cite{9746939, yi2023add2023secondaudio}, DFADD \cite{du2024dfadd}, CodecFake \cite{xie2025codecfake}, and CD-ADD \cite{li2024cross} expand coverage of synthesis and codec variations, while MLAAD \cite{muller2024mlaad}, SpeechFake \cite{huang2025speechfake}, and SpoofCeleb \cite{jung2025spoofceleb} enhance linguistic and speaker diversity. Nevertheless, distortions introduced by real social media dissemination are often absent or only partially modeled. ITW \cite{muller22_interspeech} moved evaluation closer to real-world conditions and demonstrated substantial performance degradation compared with ASVspoof2019, yet it remains monolingual (English) and limited to a single platform (YouTube), constraining broader cross-regional analysis. FSW \cite{xie2026fake} further extends evaluation across four Chinese platforms, representing an important step toward realistic benchmarking within a specific linguistic and regional context. However, a unified benchmark that encompasses diverse languages, regions, and platforms remains lacking, making it difficult to fully evaluate model generalization in practical environments.

To address this gap, we introduce ML-ITW, a multilingual in-the-wild benchmark covering 14 languages, seven major social media platforms, and 180 public figures, totaling 28.39 hours of audio. Using a unified evaluation protocol, we assess representative end-to-end models, self-supervised hybrids, and audio large language models across ASVspoof2019-LA, ITW, and ML-ITW. Results show consistent and significant performance degradation under multilingual and multi-platform conditions, highlighting persistent generalization limitations of current detection approaches in realistic dissemination environments.

\begin{samepage}
Overall, the main contributions of this work are summarized as follows:\par
\end{samepage}
\begin{itemize}
\item We introduce ML-ITW, a multilingual in-the-wild audio deepfake benchmark spanning 14 languages, seven social media platforms, and 180 public figures, reflecting real-world dissemination and providing a challenging evaluation benchmark for cross-lingual and multi-speaker generalization.
\item We conduct unified cross-dataset and cross-lingual evaluations of representative end-to-end, self-supervised hybrid, and audio large language models on ASVspoof2019-LA, ITW, and ML-ITW, demonstrating significant performance degradation and language-dependent variability under realistic multilingual conditions.
\item The experimental results show that reliance on controlled benchmarks may overestimate model robustness, and emphasize the importance of diverse, real-world evaluation settings for advancing reliable and generalizable audio deepfake detection systems.
\end{itemize}

\section{ML-ITW Dataset Construction}

\subsection{Dataset Collection}

The ML-ITW dataset was constructed from publicly available multimedia content collected from seven major social media and video platforms, including Bilibili, Facebook, TikTok, Douyin, X, Rednote, and YouTube. A total of 180 public figures, including politicians and internationally recognized celebrities, were selected as target speakers. The dataset covers 14 languages to reflect the linguistic diversity encountered in real-life deployment scenarios.

\begin{table}[t]
  \caption{Statistical overview of in-the-wild datasets compared with ML-ITW.}
  \label{tab:dataset_comparison}
  \centering
  \small 
  \setlength{\tabcolsep}{2pt} 
  \begin{tabularx}{\linewidth}{X c c c c} 
      \toprule
      \textbf{Dataset} & \textbf{\#Spk} & \textbf{Hours} & \textbf{Languages} & \textbf{Platforms} \\
    \midrule
    ITW \cite{muller22_interspeech} & 54    & 37.90   & English & 1 \\
    FSW \cite{xie2026fake}          & 128   & 254.58  & Chinese & 4 \\
    \midrule
    \textbf{ML-ITW} & \textbf{180} & \textbf{28.39} & \textbf{14 langs} & \textbf{7} \\
    \bottomrule
  \end{tabularx}
\end{table}

For each speaker, both spoofed and bona fide speech were collected. Spoofed samples were required to satisfy at least one of the following criteria:
\begin{itemize}
\item The video explicitly stated that the audio was generated or synthesized using AI technologies.
\item The spoken content was clearly inconsistent with the speaker's public identity, known stance, or contextual behavior, indicating a high likelihood of synthetic manipulation.
\end{itemize}
To reduce labeling noise, we apply a conservative validation strategy. Explicitly labeled AI-generated samples were directly included, while context-based cases were cross-checked against verified speeches and reliable sources. Ambiguous or uncertain instances were excluded from the dataset to prioritize label reliability.

For each spoofed instance, corresponding real speech recordings were subsequently retrieved from verified interviews, public speeches, or official appearances of the same individual. This paired collection strategy ensures speaker consistency while preserving realistic variability in recording conditions.

\subsection{Data Standardization and Segmentation}

\begin{table}[t]
  \caption{Distribution of the ML-ITW dataset across different social media platforms.}
  \label{tab:platform_distribution}
  \centering
  \setlength{\tabcolsep}{3pt} 
  \begin{tabular}{l c c c c c c}
    \toprule
    \multirow{2}{*}{\textbf{Platform}} & \multicolumn{2}{c}{\textbf{Real}} & \multicolumn{2}{c}{\textbf{Fake}} & \multicolumn{2}{c}{\textbf{Total}} \\
    \cmidrule(lr){2-3} \cmidrule(lr){4-5} \cmidrule(lr){6-7}
     & \textbf{\# Utt} & \textbf{Hrs} & \textbf{\# Utt} & \textbf{Hrs} & \textbf{\# Utt} & \textbf{Hrs} \\
    \midrule
    YouTube   & 5,948 & 8.00 & 531   & 0.61 & 6,479 & 8.61 \\
    Bilibili  & 4,263 & 4.57 & 793   & 0.83 & 5,056 & 5.40 \\
    Facebook  & 3,027 & 3.50 & 1,129 & 1.18 & 4,156 & 4.68 \\
    X (Twitter) & 3,046 & 2.98 & 1,061 & 1.29 & 4,107 & 4.27 \\
    Rednote   & 1,333 & 1.50 & 1,393 & 1.42 & 2,726 & 2.92 \\
    Douyin    & 182   & 0.26 & 1,149 & 1.41 & 1,331 & 1.67 \\
    TikTok    & 369   & 0.47 & 305   & 0.35 & 674   & 0.82 \\
    \midrule
    \textbf{Total} & \textbf{18,168} & \textbf{21.29} & \textbf{6,361} & \textbf{7.09} & \textbf{24,529} & \textbf{28.39} \\
    \bottomrule
  \end{tabular}
\end{table}

\begin{table*}[th]
  \caption{Performance comparison of different models on ASVspoof 2019 LA, ITW, and ML-ITW datasets.}
  \label{tab:overall_performance}
  \centering
  \setlength{\tabcolsep}{3pt}
  \begin{tabular}{l cccc c cccc c cccc}
    \toprule
    \multirow{2}{*}{\textbf{Model}} & \multicolumn{4}{c}{\textbf{ASVspoof2019-LA}} & & \multicolumn{4}{c}{\textbf{ITW}} & & \multicolumn{4}{c}{\textbf{ML-ITW}} \\
    \cmidrule{2-5} \cmidrule{7-10} \cmidrule{12-15}
    & \textbf{EER}$\downarrow$ & \textbf{AUC}$\uparrow$ & \textbf{ACC}$\uparrow$ & \textbf{F1}$\uparrow$ & & \textbf{EER}$\downarrow$ & \textbf{AUC}$\uparrow$ & \textbf{ACC}$\uparrow$ & \textbf{F1}$\uparrow$ & & \textbf{EER}$\downarrow$ & \textbf{AUC}$\uparrow$ & \textbf{ACC}$\uparrow$ & \textbf{F1}$\uparrow$ \\
    \midrule
    \rowcolor{violet!10} \multicolumn{15}{l}{\textit{\textbf{End-to-End Models}}} \\
    LCNN \cite{lavrentyeva2019stc}        & $7.89$  & $97.38$ & $92.11$ & $95.44$ & & $51.47$ & $47.63$ & $48.54$ & $41.22$ & & $48.84$ & $51.22$ & $51.16$ & $35.20$ \\
    RawNet2 \cite{tak2021end}    & $6.02$  & $97.73$ & $88.66$ & $93.25$ & & $44.90$ & $57.81$ & $55.00$ & $47.78$ & & $47.59$ & $53.76$ & $41.07$ & $40.22$ \\
    RawGAT-ST \cite{tak2021endtoendspectrotemporalgraphattention}   & $1.92$  & \cellcolor{gray!20}$\mathbf{99.83}$ & $96.44$ & $97.98$ & & $33.19$ & $73.32$ & $49.20$ & \cellcolor{gray!20}$\mathbf{57.56}$ & & $42.12$ & $58.82$ & $28.80$ & $39.05$ \\
    LibriSeVoc \cite{Sun_2023_CVPR}  & $36.55$ & $68.44$ & $29.21$ & $35.03$ & & \cellcolor{gray!20}$\mathbf{26.22}$ & \cellcolor{gray!20}$\mathbf{79.66}$ & \cellcolor{gray!20}$\mathbf{67.00}$ & $37.83$ & & $42.00$ & \cellcolor{gray!20}$\mathbf{61.42}$ & \cellcolor{gray!20}$\mathbf{66.88}$ & $35.17$ \\
    AASIST \cite{jung2022aasist}     & \cellcolor{gray!20}$\mathbf{1.89}$  & $99.71$ & \cellcolor{gray!20}$\mathbf{98.14}$ & \cellcolor{gray!20}$\mathbf{98.96}$ & & $52.13$ & $49.92$ & $43.78$ & $48.48$ & & \cellcolor{gray!20}$\mathbf{41.85}$ & $60.60$ & $26.40$ & \cellcolor{gray!20}$\mathbf{40.97}$ \\
    \midrule
    \rowcolor{indigo!10} \multicolumn{15}{l}{\textit{\textbf{Self-Supervised Representation-Based Models}}} \\
    XLSR+\allowbreak AASIST \cite{tak2022automatic} & \cellcolor{gray!20}$\mathbf{0.15}$  & \cellcolor{gray!20}$\mathbf{99.99}$ & \cellcolor{gray!20}$\mathbf{99.72}$ & \cellcolor{gray!20}$\mathbf{99.84}$ & & $7.25$  & $97.30$ & $56.03$ & $62.75$ & & $49.40$ & $51.37$ & $27.01$ & $40.79$ \\
    ML\_SSLFG \cite{tran2025multi}  & $0.45$ & $99.98$ & $99.55$ & $99.75$ & & \cellcolor{gray!20}$\mathbf{5.06}$ & \cellcolor{gray!20}$\mathbf{98.87}$ & \cellcolor{gray!20}$\mathbf{94.93}$ & \cellcolor{gray!20}$\mathbf{93.30}$ & & \cellcolor{gray!20}$\mathbf{40.49}$ & \cellcolor{gray!20}$\mathbf{59.87}$ & \cellcolor{gray!20}$\mathbf{58.19}$ & \cellcolor{gray!20}$\mathbf{41.92}$ \\
    XLSR+\allowbreak SLS \cite{zhang2024audio}   & $40.45$ & $63.30$ & $68.63$ & $80.28$ & & $47.85$ & $53.03$ & $46.02$ & $50.31$ & & $48.83$ & $51.74$ & $38.03$ & $39.68$ \\  
    \midrule
    \rowcolor{blue!10} \multicolumn{15}{l}{\textit{\textbf{Audio Large Language Models}}} \\
    ALLM4ADD \cite{gu2025allm4add}    & \cellcolor{gray!20}$\mathbf{0.43}$ & \cellcolor{gray!20}$\mathbf{99.97}$ & \cellcolor{gray!20}$\mathbf{99.39}$ & \cellcolor{gray!20}$\mathbf{99.5}$ & & $26.99$ & \cellcolor{gray!20}$\mathbf{80.83}$ & \cellcolor{gray!20}$\mathbf{79.28}$ & \cellcolor{gray!20}$\mathbf{61.57}$ & & $43.7$ & \cellcolor{gray!20}$\mathbf{67.25}$ & \cellcolor{gray!20}$\mathbf{68.98}$ & \cellcolor{gray!20}$\mathbf{45.26}$ \\ 
    HoliAntiSpoof \cite{xu2026holiantispoof}    & $1.11$ & $97.8$ & $96.59$ & $97$ & & \cellcolor{gray!20}$\mathbf{18.51}$ & $75.63$ & $70.36$ & $45.69$ & & \cellcolor{gray!20}$\mathbf{42.45}$ & $53.78$ & $57.01$ & $42.56$ \\ 
    FT-GRPO \cite{xie2026interpretable}    & $0.52$ & $99.67$ & $98.79$ & $99.33$ & & $26.97$ & $68.82$ & $52.42$ & $45.76$ & & $50.03$ & $49.97$ & $25.47$ & $40.60$ \\ 
    \bottomrule
  \end{tabular}
\end{table*}

All videos were converted to a unified audio format. Audio was extracted from MP4 files using FFmpeg, converted to mono, and resampled to 16 kHz, then saved as uncompressed WAV files. No denoising or source separation was introduced, preserving real-world acoustic characteristics. Speech segmentation employed a dual-backend VAD framework. Silero VAD \footnote{https://github.com/snakers4/silero-vad} served as the primary detector with a 0.5 threshold, 250 ms minimum speech duration, 200 ms minimum silence, and 50 ms padding. 

To highlight the distinct contributions of this work, we present a statistical comparison between ML-ITW and existing audio deepfake in-the-wild benchmarks in Table \ref{tab:dataset_comparison}. The ML-ITW dataset consists of 24,529 speech segments, totaling approximately 28.4 hours of audio, including 18,168 bona fide and 6,361 spoofed segments. The average duration per speaker is approximately 9.46 minutes, with an average utterance length of 4.17 seconds. The platform distribution is detailed in Table \ref{tab:platform_distribution}. In terms of linguistic diversity, the dataset covers 14 languages. English and Chinese constitute the primary subsets, accounting for over 19 hours combined. Japanese, German, Portuguese, and Korean provide substantial coverage (approx. 1–2 hours each), while the remaining tail includes Spanish, Turkish, Hindi, Russian, French, Hungarian, Ukrainian, and Hebrew, ensuring broad representation for cross-lingual evaluation.
\section{Experimental Setup}

\subsection{Evaluation of Three Model Categories}

To comprehensively assess the robustness and generalization of speech anti-spoofing systems on ML-ITW, we evaluate three categories of models: end-to-end architectures (LCNN, RawNet2, RawGAT-ST \cite{tak2021endtoendspectrotemporalgraphattention}, LibriSeVoc \cite{Sun_2023_CVPR}, AASIST), self-supervised representation-based models (ML\_SSLFG, XLSR+\allowbreak SLS \cite{zhang2024audio}, XLSR+\allowbreak AASIST), and audio large language models (ALLM4ADD, HoliAntiSpoof, FT-GRPO \cite{xie2026interpretable}). In addition to ML-ITW, model performance is also assessed on the ASVspoof2019-LA test set and the ITW dataset.

For all models, we use publicly available pretrained checkpoints whenever possible. If pretrained weights are not provided, the detector is trained on the ASVspoof2019-LA training set. In all cases, we adhere to the original hyperparameters and training protocols. The checkpoint with the best development-set performance is selected for final evaluation.

For consistency and efficiency, a fixed input length of 64,600 samples (approximately four seconds) is used. Longer utterances are truncated, while shorter ones are repeated to reach the target length, avoiding zero-padding to prevent introducing silence that could bias results.
\subsection{Cross-Dataset Training Evaluation}

To assess cross-dataset generalization, we perform a controlled comparison using the XLSR+\allowbreak AASIST architecture with XLSR-300M as the fixed frontend encoder. Individual models are trained on ASVspoof5, ASVspoof2019-LA, CD-ADD, Codecfake, DFADD, FSW, SpeechFake, and SpoofCeleb. Training settings remain consistent, with a batch size of 16, a learning rate of $1 \times 10^{-6}$, and a maximum of 100 epochs. Early stopping occurs if the development-set EER does not improve for 10 consecutive epochs. RawBoost data augmentation \cite{tak2022rawboost} is applied throughout training. After training on each dataset, models are evaluated on ML-ITW using identical preprocessing and protocols, enabling a fair comparison of how various training data affect cross-domain performance on real-world, multilingual deepfake speech.

\subsection{Evaluation Metrics}

To rigorously evaluate detection performance and generalization, we adopt standard metrics including Accuracy (ACC), F1-score, Area Under the ROC Curve (AUC), and Equal Error Rate (EER), with EER serving as the primary evaluation criterion.

\section{Evaluation Results}

\subsection{Overall System-Level Performance}

Table \ref{tab:overall_performance} summarizes results on ASVspoof2019-LA, ITW, and ML-ITW. All three categories of models achieve near-saturated performance on ASVspoof2019-LA, with EERs typically very low and AUCs approaching 100\%. On ITW, performance becomes more differentiated: several self-supervised and large language models retain moderate robustness, whereas traditional architectures degrade more noticeably. However, this advantage diminishes on ML-ITW. Across all model types, EERs rise to the 40–50\% range and AUC values approach near-random levels, indicating that multilingual and cross-platform distribution shifts substantially undermine learned decision boundaries.

The consistent performance degradation across architectures suggests that the challenge extends beyond model capacity. Real-world transmission effects, such as compression, re-encoding, and platform-specific processing, alter spectral structure in ways not captured by controlled benchmarks. Furthermore, exposure to a limited set of synthesis methods during training yields limited robustness to the broader and evolving generation techniques reflected in ML-ITW. Significantly, the comparable degradation indicates that neither architectural sophistication nor large-scale pretraining alone guarantees stable generalization under realistic multilingual conditions.

\begin{table*}[th]
  \caption{Impact of training data distribution on cross-dataset generalization. All systems employ the fixed \textbf{XLSR+\allowbreak AASIST} architecture but are trained on different datasets.}
  \label{tab:cross_dataset_training}
  \centering
  \setlength{\tabcolsep}{3pt}
  \begin{tabular}{l cccc c cccc c cccc}
    \toprule
    \multirow{2}{*}{\textbf{Training Set}} & \multicolumn{4}{c}{\textbf{ASVspoof2019-LA}} & & \multicolumn{4}{c}{\textbf{ITW}} & & \multicolumn{4}{c}{\textbf{ML-ITW}} \\
    \cmidrule{2-5} \cmidrule{7-10} \cmidrule{12-15}
    & \textbf{EER}$\downarrow$ & \textbf{AUC}$\uparrow$ & \textbf{ACC}$\uparrow$ & \textbf{F1}$\uparrow$ & & \textbf{EER}$\downarrow$ & \textbf{AUC}$\uparrow$ & \textbf{ACC}$\uparrow$ & \textbf{F1}$\uparrow$ & & \textbf{EER}$\downarrow$ & \textbf{AUC}$\uparrow$ & \textbf{ACC}$\uparrow$ & \textbf{F1}$\uparrow$ \\
    \midrule
    ASVspoof5 \cite{wang24_asvspoof}    & $16.91$ & $90.75$ & $28.19$ & $33.22$ & & $14.39$ & $92.16$ & $74.83$ & $49.00$ & & $42.89$ & $61.29$ & $74.90$ & $7.82$ \\
    ASVspoof2019-LA \cite{wang2020asvspoof}  & \cellcolor{gray!20}$\mathbf{0.15}$  & \cellcolor{gray!20}$\mathbf{99.99}$ & \cellcolor{gray!20}$\mathbf{99.72}$ & \cellcolor{gray!20}$\mathbf{99.84}$ & & $7.25$  & $97.30$ & $56.03$ & $62.75$ & & $49.40$ & $51.37$ & $27.01$ & $40.79$ \\
    CD-ADD \cite{li2024cross}        & $9.59$  & $96.67$ & $57.74$ & $69.23$ & & $9.53$  & $93.82$ & 87.28 & $80.02$ & & $35.12$ & $69.67$ & $75.10$ & $11.22$ \\
    Codecfake \cite{xie2025codecfake}    & $1.99$  & $99.81$ & $97.34$ & $98.50$ & & $5.06$  & $98.61$ & $86.37$ & $84.40$ & & $35.89$ & $68.19$ & $57.77$ & $47.81$ \\
    DFADD \cite{du2024dfadd}        & $16.04$ & $91.35$ & $78.00$ & $86.02$ & & $11.81$ & $94.59$ & $61.39$ & $65.40$ & & $45.28$ & $56.46$ & $27.20$ & $41.38$ \\
    FSW \cite{xie2026fake}          & $49.45$ & $53.16$ & $33.21$ & $41.92$ & & $47.98$ & $52.70$ & $62.94$ & $1.29$  & & $36.79$ & $68.50$ & $75.54$ & $14.17$ \\
    SpeechFake-BD \cite{huang2025speechfake} & $10.55$ & $96.31$ & $89.22$ & $93.69$ & & \cellcolor{gray!20}$\mathbf{2.62}$  & \cellcolor{gray!20}$\mathbf{99.42}$ & $74.21$ & $74.20$ & & $32.92$ & $72.48$ & $65.20$ & \cellcolor{gray!20}$\mathbf{51.01}$ \\
    SpoofCeleb \cite{jung2025spoofceleb}   & $20.79$ & $87.81$ & $48.01$ & $59.21$ & & $3.59$  & $98.96$ & \cellcolor{gray!20}$\mathbf{96.19}$  & \cellcolor{gray!20}$\mathbf{94.80}$ & & \cellcolor{gray!20}$\mathbf{27.65}$ & \cellcolor{gray!20}$\mathbf{79.62}$ & \cellcolor{gray!20}$\mathbf{79.73}$ & $42.41$ \\
    \bottomrule
  \end{tabular}
\end{table*}

\begin{table}[t!]
  \caption{Language-wise EER (\%)$\downarrow$ comparison on ML-ITW. The bottom two rows report the Macro Average (mean EER across all languages) and Std. Dev. (standard deviation), indicating overall performance and cross-lingual stability.}
  \label{tab:lang_eer}
  \centering
  \setlength{\tabcolsep}{3pt} 
  \begin{tabular}{l c c c}
    \toprule
    \textbf{Language} & \textbf{AASIST} & \textbf{ML\_SSLFG} & \textbf{ALLM4ADD} \\
    \midrule
    Chinese    & $51.99$ & $39.30$ & $49.46$ \\
    English    & $33.66$ & $44.99$ & $45.76$ \\
    French     & $16.40$ & $80.42$ & $42.97$ \\
    German     & $40.68$ & $52.47$ & $39.28$ \\
    Hebrew     & $45.30$ & $1.39$  & $43.81$ \\
    Hindi      & $37.38$ & $42.77$ & $46.89$ \\
    Hungarian  & $4.73$  & $33.39$ & $35.31$ \\
    Japanese   & $40.08$ & $50.17$ & $46.82$ \\
    Korean     & $47.29$ & $35.52$ & $51.95$ \\
    Portuguese & $39.83$ & $35.21$ & $49.48$\\
    Russian    & $9.10$  & $34.07$ & $47.33$\\
    Spanish    & $34.97$ & $66.26$ & $37.66$\\
    Turkish    & $38.09$ & $42.02$ & $40.42$\\
    Ukrainian  & $53.87$ & $58.86$ & $50.06$\\
    \midrule
    \textbf{Macro Average} & $\mathbf{35.24}$ & $44.06$ & $44.80$ \\
    \textbf{Std. Dev.}     & $15.03$ & $18.73$ & $\mathbf{4.87}$ \\
    \bottomrule
  \end{tabular}
  
\end{table}

\subsection{Impact of Training Data Distribution on Cross-Dataset Generalization}

As shown in Table \ref{tab:cross_dataset_training}, models trained on ASVspoof2019-LA achieve near-saturated performance on its in-domain test set, but this advantage does not transfer to more realistic conditions. On ITW, systems trained on SpeechFake, SpoofCeleb, and Codecfake outperform the ASVspoof2019-trained model, indicating that strong performance on a controlled benchmark does not necessarily translate to better generalization when the evaluation distribution changes. The gap widens further on ML-ITW, where all models experience substantial degradation, and EERs rise to 30\%–50\%. Although training on SpoofCeleb and SpeechFake leads to comparatively lower error rates than training on ASVspoof2019-LA, none of the training sets maintain stable performance in the setting of ML-ITW. Since the model architecture, optimization strategy, and preprocessing pipeline are strictly controlled, the observed performance differences can be attributed to variations in training data distribution. This comparison demonstrates that models remain highly sensitive to unseen linguistic, acoustic, and propagation variations, even when trained on diverse corpora. Datasets that partially reflect real-world distortions offer modest improvements, but these gains are still insufficient to address the broader linguistic and propagation diversity, confirming that existing corpora lack the coverage necessary for stable generalization in realistic dissemination scenarios.

\subsection{Language-wise Analysis}

To ensure representativeness, we select AASIST, ML\_SSLFG and ALLM4ADD, which demonstrate strong performance in their respective categories, enabling an architecture-focused comparison at comparable performance levels. Table \ref{tab:lang_eer} shows their performance across languages on ML-ITW. There is no consistent difficulty ranking across languages, yet distinct error patterns still emerge. AASIST achieves the lowest macro-average EER (35.24\%), though it still exhibits substantial variation across languages. ML\_SSLFG yields a higher average error rate (44.06\%) and displays extreme imbalances in certain languages, with some results approaching random performance. In contrast, ALLM4ADD attains a macro-average EER of 44.80\%, which is the highest among the three, but its standard deviation is only 4.87, indicating the most concentrated error distribution. The three models also differ in their ranking of language difficulty, suggesting that performance variation arises not only from language characteristics but also from differences in how each architecture encodes language-specific acoustic features and spoofing cues.

This variation might arise from distinct representation learning paradigms. AASIST employs end-to-end optimization, directly aligning features with the spoof detection objective to learn stable task-specific cues. ML\_SSLFG relies on self-supervised pretraining followed by task adaptation, which may leave language statistics and spoofing artifacts insufficiently disentangled under homogeneous training distributions. In contrast, ALLM4ADD leverages large-scale contextual modeling, which tends to smooth language-dependent fluctuations and yields more uniform behavior across languages. Results thus reveal differing stability profiles across modeling paradigms and highlight ongoing limitations in achieving both cross-lingual consistency and high discriminative power in real-world multilingual conditions.
\section{Discussion}

Driven by rapid advances in speech synthesis and ongoing updates in platform-level encoding and compression, we construct ML-ITW to encompass a wide range of languages, speakers, and social media sources, thereby providing a more comprehensive and realistic evaluation benchmark for the evaluation of deepfake detection systems. Model accuracy declines significantly on ML-ITW, revealing major generalization gaps under multilingual, multi-speaker, and cross-platform conditions. These findings suggest that modern forgery artifacts and real-world transmission effects appear to disrupt decision boundaries established on controlled datasets.

In addition to quantifying performance loss, this work provides empirical evidence for reexamining evaluation standards in audio deepfake detection. ML-ITW serves as a more realistic benchmark for assessing cross-lingual and cross-platform robustness, highlighting the need to incorporate propagation variability into both dataset design and model development. One remaining limitation is the relatively small sample size for some low-resource languages, which may affect the reliability of language-level analyses. Future efforts will seek to improve language balance, enhancing both statistical reliability and cross-lingual evaluation. In parallel, expanding both the quantity and diversity of audio sources will better reflect the evolving landscape of online speech synthesis and dissemination, and further strengthen the validity of benchmark evaluations.

\section{Conclusion}
This paper introduces ML-ITW, a multilingual in-the-wild benchmark for speech deepfake detection, comprising bona fide and spoofed speech collected from seven social media platforms across 14 languages and 180 public figures. Systematic evaluations of end-to-end models, self-supervised representation-based models, and audio large language models show that performance advantages on conventional benchmarks do not readily transfer to ML-ITW. Cross-dataset training experiments further demonstrate that the choice of training data can substantially improve detection performance but remains insufficient to close the generalization gap in real-world conditions. Language-wise analyses also reveal that detection ability and cross-lingual stability do not consistently align across models. These findings highlight the need to consider training data, language coverage, and real-world dissemination conditions jointly when evaluating speech deepfake detectors. ML-ITW provides a publicly available benchmark for such evaluations. Future work will expand the dataset and improve language balance to support more comprehensive studies of generalization.
\clearpage

\bibliographystyle{IEEEtran}
\begingroup
\setlength{\baselineskip}{10pt}
\bibliography{mybib}
\endgroup
\end{document}